\newcommand{\CLASSINPUTtoptextmargin}{1in}
\newcommand{\CLASSINPUTbottomtextmargin}{0.75in}

\documentclass[conference]{IEEEtran}
\IEEEoverridecommandlockouts
\usepackage{amsmath,amssymb,amsfonts}
\usepackage{algorithmic}
\usepackage{graphicx}
\usepackage[caption=false, font=footnotesize]{subfig}
\usepackage{textcomp}
\usepackage{threeparttable}
\usepackage{multirow}
\usepackage[table,xcdraw]{xcolor}
\usepackage{svg}
\usepackage{calc}

\renewcommand{\baselinestretch}{0.992}

\newcommand{\eegtcnet}[0]{\textsc{EEG-TCNet}}

\def\BibTeX{{\rm B\kern-.05em{\sc i\kern-.025em b}\kern-.08em
    T\kern-.1667em\lower.7ex\hbox{E}\kern-.125emX}}
\begin{document}

\title{EEG-TCNet: An Accurate Temporal Convolutional Network for Embedded Motor-Imagery Brain--Machine Interfaces
%EEG-TCNet: A Compact and Flexible TCN for Accurate Motor-Imagery Brain--Machine Interfaces
}
% We need another title which distinguishes from the one of EEGNet: "EEGNet: a compact convolutional neural network for EEG-based brain–computer interfaces

% \author{
% \IEEEauthorblockN{Authors removed for review.}
% \IEEEauthorblockA{}
% }

\author{\IEEEauthorblockN{
    Thorir Mar Ingolfsson\IEEEauthorrefmark{1}, 
    Michael Hersche\IEEEauthorrefmark{1}, 
    Xiaying Wang\IEEEauthorrefmark{1},
    Nobuaki Kobayashi\IEEEauthorrefmark{2},
    Lukas Cavigelli\IEEEauthorrefmark{1}\IEEEauthorrefmark{3},
    Luca Benini\IEEEauthorrefmark{1}}
    % \IEEEauthorblockA{\\[-2mm]\IEEEauthorrefmark{1}ETH Zürich, Dept. EE \& IT,  Switzerland \hspace{15mm}
    % \IEEEauthorrefmark{2}Huawei Technologies, Zurich Research Center, Switzerland \\
    % \IEEEauthorrefmark{3}University of Bologna, DEI, Italy}
    \IEEEauthorblockA{\\[-2mm]\IEEEauthorrefmark{1}ETH Zürich, Dept. EE \& IT,  Switzerland \hspace{15mm}\IEEEauthorrefmark{2}Nihon University, College of Science and Technology, Japan   \\
    \IEEEauthorrefmark{3}Huawei Technologies, Zurich Research Center, Switzerland\vspace{-0.3cm}}
    %\IEEEauthorblockA{\{xiaywang, herschmi, magnom, lbenini\}@iis.ee.ethz.ch}
    \thanks{Corresponding emails: \{thoriri, herschmi, xiaywang\}@iis.ee.ethz.ch}
   % \thanks{Corresponding email: thoriri@ethz.ch}
    }

% \author{\IEEEauthorblockN{1\textsuperscript{st} Thorir Mar Ingolfsson}
% \IEEEauthorblockA{\textit{Integrated Systems Laboratory} \\
% \textit{ETH Zürich}\\
% Zürich, Switzerland \\
% thoriri@student.ethz.ch}
% \and
% \and
% \and
% \and
% \and
\maketitle
\begin{abstract}
    %This paper presents a compact and flexible novel model, \emph{EEG-TCNet}, for accurate motor-imagery brain--computer interfaces (MI-BCI) classification tasks.
    In recent years, deep learning (DL) %research 
    has contributed significantly to the improvement of motor-imagery brain--machine interfaces (MI-BMIs) based on electroencephalography (EEG).
    While achieving high classification accuracy, DL models have also grown in size, requiring a vast amount of memory and computational resources.
    This poses a major challenge to an embedded BMI solution that guarantees user privacy, reduced latency, and low power consumption by processing the data locally.
    % since the data is processed locally instead of being transmitted to a remote host.
    %
    In this paper, we propose \eegtcnet{}, a novel temporal convolutional network (TCN) that 
    %is able to attain robust classification accuracy, while at the same time being compact with few trainable parameters.
    %
    achieves outstanding accuracy while requiring few trainable parameters. 
    Its low memory footprint and low computational complexity for inference make it suitable for embedded classification on resource-limited devices at the edge. 
    Experimental results on the BCI Competition IV-2a dataset show that \eegtcnet{} achieves 77.35\% classification accuracy in 4-class MI.
    %, outperforming state-of-the-art (SoA) neural networks feasible for embedded implementation. 
    %
    By finding the optimal network hyperparameters per subject, we further improve the accuracy to 83.84\%. 
    Finally, we demonstrate the versatility of \eegtcnet{} on the Mother of All BCI Benchmarks (MOABB), a large scale test benchmark containing 12 different EEG datasets with MI experiments.
    The results indicate that \mbox{\eegtcnet{}} successfully generalizes beyond one single dataset, outperforming the current state-of-the-art (SoA) on MOABB by a meta-effect of 0.25.
    % ~\cite{jayaram_moabb_2018}
    %Our proposed model performs better than all other pipelines proposed in the original MOABB paper and therefore becomes the new SoA for MOABB. 
    %
    %By looking at the final meta effect our proposed model performs better than the old SoA by a final meta effect of 0.53. The significance of this experiment is that \emph{EEG-TCNet} can generalize well outside the single BCI dataset it was modeled on.
    %
    %Even though \emph{EEG-TCNet} is modeled after a single dataset, it is still representative under the very robust validation framework that MOABB provides.
\end{abstract}


\section{Introduction}
% Intro BCI
Brain--machine interfaces (BMIs) allow direct communication between humans and external devices by analyzing neural activity of the human brain, typically recorded with noninvasive electroencephalography (EEG)~\cite{graimann_braincomputer_2010_short}.
One promising approach is based on motor-imagery (MI), which is the cognitive process of thinking about the motion of a body part, e.g., the left hand, without actually performing it.
%
%MI-EEG are EEG signals of imagining body movement without actual movement. 
%
%MI-BCI strives to decode that cognitive process, e.g., thinking of moving the tongue, feet, or any other body part. 
%
MI-BMIs assist people with impairments to regain independence, e.g., by steering a wheelchair~\cite{xiong_low-cost_2019_short}, controlling a prosthesis~\cite{condori_embedded_2016, cho_classification_2019}, or by enabling motor rehabilitation~\cite{cho_motor_2018}. 
%
%Recently, MI-BMIs have gained attention beyond medical applications, e.g., in gaming~\cite{bonnet_two_2013} or augmented reality~\cite{}. 
% Challenges in BCI

However, successful decoding of MI-based EEG signals remains a challenging task, mainly due to a low signal-to-noise ratio and high variance among different subjects, which prohibits the use of a single MI-BMI model for all subjects~\cite{lotte_review_2018}.
% 
%On the other hand, doing long calibration measurements can be cumbersome and tiring, resulting in a limited amount of labeled training data. 
% CNNs in MI-BMI
%Therefore, 
Conventional approaches rely on domain-specific knowledge, mostly using handcrafted feature extractors, such as filter bank common spatial pattern (FBCSP)~\cite{ang_filter_2008} or Riemannian covariance~\cite{hersche_fast_2018} features in combination with robust classifiers like linear discriminant analysis (LDA) or support vector machines (SVMs)~\cite{lotte_review_2018}. 

% The rise of CNNs
Recently, convolutional neural networks (CNNs) have gained increasing attention in the MI-BMI field, reducing the data pre-processing steps and eliminating the procedure of handcrafting features.  
%, since they allow an end-to-end trainable architecture. 
%
One of the first successful CNN in MI classification was Shallow ConvNet~\cite{schirrmeister_deep_2017}, which was inspired by FBCSP.
% and achieved state-of-the-art (SoA) accuracy on the 4-class BCI competition IV-2a dataset. 
%
The more compact and generally applicable EEGNet~\cite{lawhern_eegnet_2018}, as well as more complex and accurate models~\cite{li_densely_2019,zhao_improvement_2017}, have extended the landscape of CNNs in MI classification. 
The most complex network is TPCT~\cite{li_novel_2020}, which achieves the state-of-the-art (SoA) accuracy of 88.87\% on the 4-class MI BCI Competition IV-2a dataset~\cite{Brunner2008BCIA}.

% Embedded MI-BCI on the edge
These networks are commonly deployed on desktop platforms or cloud servers, however, running MI classification on  remote computers raises serious concerns in terms of latency, availability, and privacy~\cite{Landau2020, 8661604, 8668446}. 
%which are unsuitable for battery-operated mobile devices and interactive applications due to their weaknesses in power consumption, latency, availability, and privacy. 
%
Processing the data near the sensor on a low-power microcontroller unit (MCU) allows us to mitigate these concerns.
% Challenges in embedding
However, accurate networks such as the TPCT model have 7.78\,M trainable parameters and require 1.73\, billion multiply-accumulate (MAC) operations per inference, which is out of reach of a typical low-power MCU with few MB of Flash and few hundreds of kB of RAM~\cite{wang_fann--mcu_2020}. 
Alternatively, more compact models such as EEGNet with 2.5\,k parameters and 13\,MMACs can come to the rescue and have been successfully implemented on MCUs~\cite{wang_accurate_2020, schneider_2020}. 
Still, they come at the cost of significantly lower classification accuracy of 72.40\%. 
%One can quantize, reduce the model~\cite{wang_accurate_2020} [cite Wang MEMEA2020], or use novel BCI-specific compression algorithms~\cite{hersche_compressing_2020} [cite Hersche DATE2020].
% 
A model that combines the best of both worlds (i.e., compactness at high accuracy) is highly desirable. 
% TCN as a light-weight alternative 
%

One viable option to boost the performance in accuracy is to use temporal convolutional networks (TCNs), which are achieving SoA accuracy on many time series classification and modeling tasks~\cite{bai_empirical_2018,lu_deep_2020}.
TCNs are capable to exponentially extend their receptive field size with only a linear increase in the number of parameters and number of MACs, unlike traditional CNNs, which show only a linear increase in the receptive field size. 
%
%This is noteworthy as traditional CNN architectures are only able to increase their receptive field size linearly with the number of parameters and  MACs.
% 
Moreover, in contrast to other time series classification networks, like recurrent neural networks (RNNs), TCNs do not suffer from exploding or vanishing gradient issues particularly, when training on long input sequences. 
%
%Another significant part about TCNs is their ability to be parallelized because convolutions can be done in parallel since the same kernel size is used in each layer. 
%
%Therefore, in both training and evaluation, a lengthy input time series can be processed as a whole in the TCN. 
%
%The TCN also has a backpropagation path different from the temporal direction of the time series, which allows TCN to avoid the problem of exploding/vanishing gradients, which plague many recurrent architectures. 
%This also results in low memory requirement for training as the kernel sizes are shared across a layer, with the backpropagation path depending only on network depth.

% Here contributions
In this paper, we introduce \eegtcnet{}, which features both the compactness of EEGNet and the high accuracy of TCNs. 
%
%It requires only 4336 trainable parameters and 13 MMACs per inference, making it suitable for resource-limited embedded devices operating at the edge. 
% 
The main contributions of the paper are as follows: 
\begin{itemize}
    \item We propose \eegtcnet{}, which 
    % is the first TCN based neural network used for MI classification. It 
    requires only 4272 trainable parameters and 6.8 MMACs per inference, making it suitable for resource-limited embedded devices. 
    % operating at the edge.
    \item We evaluate the proposed \eegtcnet{} on the BCI Competition IV-2a dataset~\cite{Brunner2008BCIA}, where it achieves a high accuracy of 77.35\%. 
    Our model requires significantly fewer parameters, MACs, and memory during inference compared to other networks with similar accuracy. 
    \item We further improve \eegtcnet{} by 6.49\% classification accuracy, reaching 83.84\%, % over 4 MI classes, 
    by finding the optimal network hyperparameters per subject based on a grid search in cross-validation on the training data. This outperforms most of the current SoA networks. % found in literature. 
    We then analyze accuracy vs. parameter counts and accuracy vs. number of MAC operations for our proposed model, and obtain that \eegtcnet{} achieves Pareto optimality in both cases.
    \item We extensively benchmark our methods on the Mother of All BCI Benchmarks (MOABB)~\cite{jayaram_moabb_2018}, where \mbox{\eegtcnet{}} outperforms the current SoA by a meta-effect of 0.25. To best of our knowledge, this work is the first external submission to MOABB, which will pave the way for reproducible results required to benchmark new BMI classifiers reliably.
\end{itemize}
Our code and the trained models are available online under a permissible open-source license for reproducibility\footnote{Will be released on GitHub after review.}.

\section{Background}
%This section introduces the BCI competition IV-2a dataset, standard models used for accurate MI classification, and temporal convolutional networks.
\subsection{BCI Competition IV-2a Dataset}
\begin{figure}
    \fontsize{8}{10}\selectfont
    \centering
    \includesvg[width=\columnwidth]{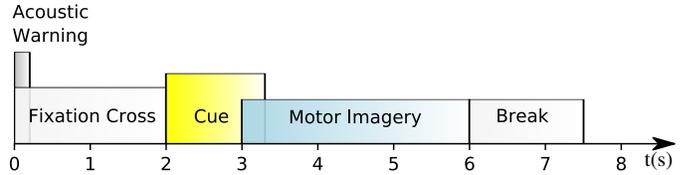}
    \caption{Timing scheme of the BCI Competition IV-2a~\cite{Brunner2008BCIA}.}
    \label{fig:timing}
\end{figure}

The BCI Competition IV-2a dataset~\cite{Brunner2008BCIA} consists of recordings from nine different subjects using 22 EEG electrodes. 
The data was collected by bandpass filtering the signals between 0.5\,Hz and 100\,Hz and sampling them at 250\,Hz. 
All subjects were requested to perform imagined movements of four different body parts: left hand, right hand, both feet, and tongue.
Besides, three electrooculography (EOG) channels give information about eye movements. 
The dataset consists of two sessions per subject recorded on different days, where we use one for training and the other for testing, and each session contains 288 trials. 
Trials containing artifacts (9.41\% of the data) were excluded from the dataset after an expert marked them based on EOG data. 
In order to go by the rules of the BCI Competition IV-2a, we make no further use of the EOG data. 
Each trial lasted 7.5\,s and was recorded according to the timing scheme shown in Fig~\ref{fig:timing}.
%fixation cross (t=0\,s)$\,\to\,$cue (t=2\,s)$\,\to\,$MI (t=3\,s)$\,\to\,$break (t=6\,s). 
%
%Note the Motor imagery part of the series is only 3 seconds long. 

\subsection{Mother of All BCI Benchmarks}\label{chap:MOABB}
The Mother of All BCI Benchmarks (MOABB)~\cite{jayaram_moabb_2018} is an aggregation of many publicly available EEG datasets, converted to a common format, and bundled in a software package. 
It was initiated because of several problems currently present in the BMI research community. 
One of those is that while many BMI datasets are made freely available, researchers do not publish code, and reproducing results required to benchmark new algorithms turns out to be more tricky than it should be. 
Moreover, performance can be significantly impacted by parameters of the pre-processing steps, toolboxes used, and implementation tricks that are rarely reported in the literature.

MOABB aims to provide solutions to these problems by building a comprehensive benchmark of popular BMI algorithms applied on an extensive list of freely available EEG datasets. 
The code is available online on GitHub; algorithms can be ranked and promoted on a website, providing a clear picture of the different solutions available in the field.
MOABB provides a variety of different datasets, both for MI and event-related potential (ERP) classification. 
In this paper, we consider all MI datasets consisting of 2--4 classes MI experiments. 
Due to internal errors in the current MOABB package, we had to replace the dataset of Zhou \textit{et al}. 2016 for the high-gamma dataset described in~\cite{schirrmeister_deep_2017}. 
The used MI datasets are summarized in Table~\ref{tab:MOABB_datasets}.
\begin{table}
    \caption{MOABB datasets attributes.}
    \label{tab:MOABB_datasets}
    \resizebox{\columnwidth}{!}{%
    \centering
    \begin{threeparttable}
        \begin{tabular}{rrrrrrr}
            \hline
            \multicolumn{1}{l}{Name} & \multicolumn{1}{l}{$N$} & \multicolumn{1}{r}{$C$} & \multicolumn{1}{l}{\# Trials} & \multicolumn{1}{r}{S} & \multicolumn{1}{l}{\# Subjects} & \multicolumn{1}{l}{Epoch (s)}\\ \hline \hline
            \rowcolor[HTML]{EFEFEF} 
            \multicolumn{1}{l}{Cho \textit{et al}} & 2 & 64 & 200 & 1 & 52 & 0-3  \\
            \multicolumn{1}{l}{Physionet} & 2 & 64 & 40-60 & 1 & 109 & 1-3 \\
            \rowcolor[HTML]{EFEFEF} 
            \multicolumn{1}{l}{Shin \textit{et al}} & 2 & 25 & 60 & 3 & 29 & 0-10 \\
            \multicolumn{1}{l}{BNCI 2014-001} & 4 & 22 & 288  & 2 & 9 & 2-6  \\
            \rowcolor[HTML]{EFEFEF} 
            \multicolumn{1}{l}{BNCI 2014-002} & 2 & 15 & 160 & 1 & 14 & 3-8\\
            \multicolumn{1}{l}{BNCI 2014-004} & 2 & 3 & 120-160 & 5 & 9 & 3-7.5 \\
            \rowcolor[HTML]{EFEFEF} 
            \multicolumn{1}{l}{BNCI 2015-001} & 2 & 13 & 200 & 2/3 & 13 & 3-8 \\
            \multicolumn{1}{l}{BNCI 2015-004} & 2 & 30 & 70-80 & 2 & 10 & 3-10 \\
            \rowcolor[HTML]{EFEFEF} 
            \multicolumn{1}{l}{Alexandre} & 2 & 16 & 40 & 1 & 9 & 0-3 \\
            \multicolumn{1}{l}{Yi \textit{et al}}  & 4 & 60 & 160 & 1 & 10 & 3-7 \\
            \rowcolor[HTML]{EFEFEF} 
            \multicolumn{1}{l}{Grosse-Wentrup \textit{et al}} & 2 & 128 & 300 & 1 & 10 & 3-10 \\
            \multicolumn{1}{l}{Schirrmeister \textit{et al}} & 4 & 128 & 260 & 1 & 14 & 0-4\\ \hline \hline
            Total: &  &  &  &  & 288 &  \\ \hline
        \end{tabular}
        \begin{tablenotes}\footnotesize
\item[] $N$ = number of classes, $C$ = number of EEG channels, $S$ = number of sessions.
\end{tablenotes}
    \end{threeparttable}
    }
\end{table} 

%%%%%%%%%%%%%%%%%%%%%%%%%%%%%%%%%%%%%%%%%%%%%%%%

\subsection{Related Work}
The BCI Competition IV-2a submissions~\cite{tangermann_review_2012} are various, but one frequent feature used by many submissions was common spatial patterns (CSP) on bandpass filtered data. 
Ang et al.~\cite{ang_filter_2008}, the winners of the competition, proposed filter bank common spatial pattern (FBCSP), which enhanced the performance of the original CSP algorithm and achieved 67.75\% classification accuracy.
%
%Other submissions then used these FBCSP features and classified them with a large variety of models, ranging from support vector machines (SVMs)~\cite{cortes_support-vector_1995} and linear discriminant analysis (LDA)~\cite{xanthopoulos_linear_2013} to other Bayesian methods. 
%
After the competition, a linear support vector machine (SVM) on Riemannian covariance matrices~\cite{hersche_fast_2018} has achieved an accuracy of 75.74\%. 
More recent studies~\cite{li_avoiding_2019} have shown that FBCSP combined with highway networks, random forests, and multiple binary classifiers can further enhance the accuracy to 78.00\%, 80.00\%, and 81.02\%, respectively. %~\cite{srivastava_highway_2015} % ~\cite{breiman_random_2001}

We categorize deep learning-based MI classification into two classes: feature input (FI) networks and raw signal input (RSI) networks. 
%We first go over RSI-networks.
The latter combine and train feature extraction and classification processes simultaneously. 
In particular, CNN-based RSI networks have been shown to achieve excellent results in MI-BMIs~\cite{lotte_review_2018}. 
EEGNet~\cite{lawhern_eegnet_2018} and Shallow ConvNet~\cite{schirrmeister_deep_2017} have achieved high accuracy with a relatively small network size; EEGNet has 1716 trainable parameters at 66.70\% accuracy, and Shallow ConvNet has 47\,324 parameters at 74.31\%. 
The main difference between the two architectures is that \mbox{Shallow ConvNet} was explicitly designed for oscillatory signal classification and hence utilizes log-band power to extract features, making it ill-suited for other similar tasks such as event-related potential (ERP) classification. 
In contrast, EEGNet is not only applicable to MI classification but also ERP tasks. 
%; for example, the P300 dataset described in~\cite{marathe_improved_2016}. -- not relevant here
%
By changing the pooling layers and expanding the network to 2036 trainable parameters, the accuracy of EEGNet has been increased to 72.40\%~\cite{uran_applying_2019}. 

Another competitive RSI-network is the MSFBCNN~\cite{wu_parallel_2019}, which utilizes multi-scale temporal convolution to extract features and achieves an accuracy of 75.80\%. 
% which is a CNN based architecture, 
%
CNN++~\cite{zhao_improvement_2017} uses not only the 22 EEG channels but also the 3 EOG channels, which in the original competition was strictly forbidden.
Inspired by CSP, CNN++ starts with a linear layer applied to each time sample, expanding the 25 channels to 30. It is followed by a CNN, which finally achieves an accuracy of 81.1\%. 
%
%These four models are considered an end-to-end approach since no feature extraction process is done beforehand.

In FI-networks, the MI classification is achieved in two stages. 
First, features are extracted from EEG signals with various approaches (e.g., CSP, spectrograms, or wavelets), and then fed into a classifier model.
DFFN~\cite{li_densely_2019} makes use of CSP to extract unique spatial filters. 
Temporal log-power features of the spatially filtered signal are then fed into a CNN that considers the correlation between adjacent layers and cross-layer features. 
This architecture closely resembles \emph{Dense Net}~\cite{huang_densely_2018}.
Network architecture hyperparameters were altered for each subject separately, achieving an accuracy of 79.71\%. 

TPCT~\cite{li_novel_2020} uses the information of electrode locations to improve classification results. 
The MI time frame was divided into 10 time windows and three sub-bands.
% covering the $\mu$ and $\beta$ bands~\cite{van_der_lubbe_lateralized_2013}. 
%
Then, for each channel, the Fast Fourier Transform (FFT) was employed to transform each time window to a spectrum, and its inverse FFT was calculated for each sub-band. 
The time-domain power features of the 10 time windows are then averaged for the same sub-band. Hence, three average power features are generated as the time-frequency features of each electrode and fed into the Clough-Tocher interpolation algorithm to generate an image with electrode information that was then classified with a VGG-like CNN. 
% ~\cite{noauthor_trivariate_nodate}
TPCT achieved an accuracy of 88.87\%, at the cost of 7.78\,M parameters and 1.73\,GMACs per inference.

\subsection{Temporal Convolutional Networks}
\begin{figure}
    \fontsize{8}{10}\selectfont
    \centering
    \includesvg[width=\columnwidth]{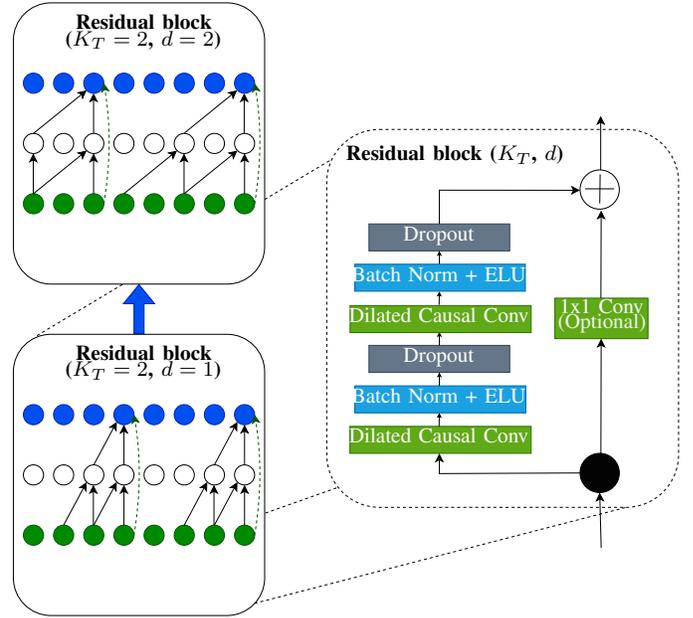}
    \caption{Architectural elements in a TCN. Left: Stacking of two residual blocks highlighting the dilated convolutions with kernel size $K_T=2$ and dilation $d=\lbrace1,2\rbrace$. Right: Detailed layers in TCN residual block. %used in the paper.
    }
    \label{fig:resblocks}
\end{figure}

In the following, we describe the generic architecture concept known as the temporal convolutional network (TCN)~\cite{bai_empirical_2018}, depicted in Fig.~\ref{fig:resblocks}. 
%
%, which actually were not the first to coin the term TCN as it had been used before in~\cite{lea_temporal_2016}.
%
%However, rather it is used as a simple descriptive term for a family of architectures. 
%
Three properties distinguish TCNs from conventional CNNs:
%1) causal convolutions; 2) dilated convolutions; 3) stacking of residual blocks.

\subsubsection{Causal Convolutions}
TCNs produce an output of the same length as the input.
%, such that there is no information flow from the future to the past. 
%
To this end, TCNs use a 1D fully-convolutional network (FCN) architecture~\cite{long_fully_2015}, where each hidden layer is the same size as the input layer, and zero-padding of length (kernel size - 1) is added to keep subsequent layers the same length as the previous ones. 
Further, causal convolutions are used to force no information flow from the future to the past. Simply put, the output at time \textit{t} depends only on inputs from time \textit{t} and earlier.

\subsubsection{Dilated Convolutions}
A regular causal convolution is only able to increase its receptive field size linearly in the depth of the network. 
This is a major disadvantage since either an extremely deep network or one with a huge kernel size is needed to obtain a large receptive field size. 
To combat this problem, TCNs use a sequence of dilated convolutions~\cite{oord_wavenet_2016}, which allows the network to increase its receptive field exponentially in size proportional to the network depth by employing a scheme of exponentially increasing dilation factors $d$.
%\begin{figure}
%\centerline{\includegraphics[scale=0.25]{DilutedConv.png}}
%\caption{Visualization of dilated causal convolutions.}
%\label{fig:dilation}
%\end{figure}
%Fig.~\ref{fig:dilation} illustrates a dilated causal convolution for three different dilation factors. When the dilation is set to $d=1$, the dilated convolution reduces to a regular casual convolution. By increasing the dilation, the top-level output can, therefore, have a much larger receptive field than with regular convolutions.
\begin{figure*}
    \fontsize{8}{10}\selectfont
    \centering
    \includesvg[width=\textwidth]{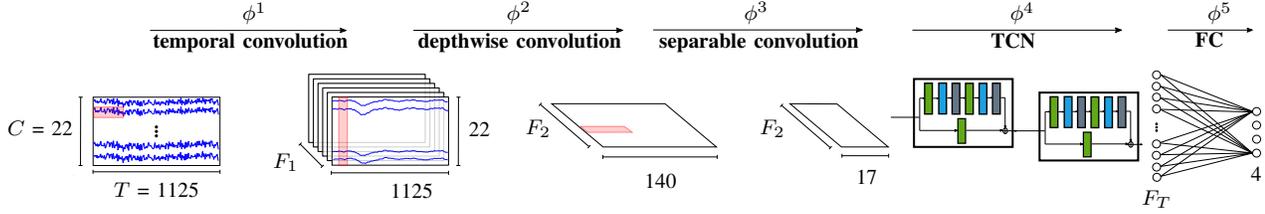}
    \caption{Architecture of the \eegtcnet{}. Where $C$ = number of EEG channels, $T$ = number of timepoints, $F_1$ = number of temporal filters, $F_2$ = number of pointwise filters and $F_T$ = number of filters in TCN module.}
    \label{fig:EEG-TCNet}
\end{figure*}

\subsubsection{Residual Blocks}
The residual block of a TCN consists of two layers of dilated convolutions, with batch normalization, non-linearity, and a dropout layer in-between the convolutions. 
Even though TCNs feature only 1D convolutions, they are still capable of processing 2D feature maps by considering the second dimension as the depth dimension.
The skip connection adds the input to the output feature map, with the check that if the depth of the input and output is different, a 1x1 convolution is put in place.
%to ensure the tensors are of the same shape. 
%
See Fig.~\ref{fig:resblocks} for an illustration of a residual block and the stacking of two residual blocks together.

By stacking residual blocks, the receptive field size increases exponentially with each residual block, as the dilation in each subsequent block is exponentially larger. 
%
%This gives us a nifty formula to calculate the receptive field size (RFS) of the TCN:
The receptive field size ($\mathrm{RFS}$) of the TCN is determined by
\begin{align}
    \mathrm{RFS} = 1 + 2 \cdot (\textit{$K_T$} - 1)\cdot(2^{\textit{L}} - 1),
\end{align}
where $K_T$ is the kernel size and $L$ the number of residual blocks.

The TCN described here slightly differs from the one explained in~\cite{bai_empirical_2018}, in the residual block in the following ways:

\begin{itemize}
    \item Batch normalization is used between convolutions instead of weight normalization as batch normalization has been shown to give higher accuracy than weight normalization on various large scale networks~\cite{gitman_comparison_2017}.
    \item We use the exponential linear unit (ELU) activation instead of the rectified linear unit (ReLU). This was done since \eegtcnet{} showed better performance with a ELU activation function than ReLU. %~\cite{nair_rectified_2010}.
    \item Instead of spatial dropout, normal dropout is used. As the TCN is applied after various convolutions the adjacent frames withing feature maps are not strongly correlated, and therefore it is beneficial to drop individual elements instead of entire 1D feature maps to regularize the activations.
\end{itemize}

\section{Methodology}
In this section, we present the main contribution of the paper. 
We show how to combine the shallow, yet discriminative feature extraction layers of EEGNet with a TCN, making use of the temporal information present in the features, which would be ignored otherwise. 
We introduce our model, named \eegtcnet{}, and analyze it either with a \textit{fixed} set of hyperparameters for all subjects, or \textit{variable} optimal subject-specific hyperparameters.

%
%Our model, named \emph{EEG-TCNet}, is further personalized by individually finding a set of optimal subject-specific hyperparameters, based on cross-validation on the training set.
% 
% 
%In the following, we describe EEG-TCNet, which extends EEGNet with a TCN to improve the processing of temporal information in MI data.
% 
%

\subsection{Data Pre-processing}
The time frame used for both training and inference on EEG data of the BCI Competition IV-2a is 0.5 seconds before the MI-cue until the end of the MI~\cite{schirrmeister_deep_2017}, resulting in a time series of 4.5\,s length, or 1125 samples. 
%
%The sample rate was kept as it was in the original dataset, 250 Hz, and no additional bandpass filtering was done. 
The sampling rate is kept at 250\,Hz, and no additional bandpass filtering is applied. 
%For some models and subjects, we observed that applying data normalization improved the performance of the TCN models. 
%
Optionally, we apply standardization by removing mean and scaling to unit-variance per channel, based on the statistics of the training set. 
%
%As stated before, trials containing artifacts are removed from the training and test set.

\subsection{\eegtcnet{}}
%We call our model EEG-TCNet and 
Fig.~\ref{fig:EEG-TCNet} illustrates the architecture of \eegtcnet{} with a more detailed description of the layers in Table~\ref{tab:EEG-TCNet_Table}. 
% show the general structure of the EEG-TCNet architecture.
%EEGNet part of EEG-TCNet
The network is, in part, inspired by the EEGNet architecture~\cite{lawhern_eegnet_2018}. 
The network starts with a 2D temporal convolution to learn frequency filters, then uses a depthwise convolution to learn frequency-specific spatial filters. 
The separable convolution learns a temporal summary for each feature map individually and then mixes the feature maps. 
The output feature map of the separable convolution still contains temporal information; therefore, the addition of a TCN further exploits temporal information. % that endured the first three layers.

%TCN part of EEG-TCNet
The first TCN block expands the $F_2$ feature maps after the separable convolution to $F_T$ feature maps. 
Overall, we stack $L$ residual blocks and select the receptive field size such that $\mathrm{RFS}\geq$ 17, allowing the TCNs to capture all temporal information available. 
% still enduring within the network. 
Finally, the last time steps of each of the $F_T$ feature maps of the last residual block are read out and fed to a fully-connected layer for classification.

\begin{table}
\caption{\eegtcnet{} architecture}
\label{tab:EEG-TCNet_Table}
%\resizebox{\columnwidth}{!}{%
\begin{threeparttable}
\centering

\begin{tabular}{lllll}
\textit{\textbf{Layer}} & \textit{\textbf{Type}} & \textit{\textbf{\#Filters}} & \multicolumn{1}{c}{\textit{\textbf{Kernel}}} & \textit{\textbf{Output}} \\ \hline
\multirow{3}{*}{$\phi^1$} & \multicolumn{1}{|l|}{Input} & \multicolumn{1}{l|}{} & \multicolumn{1}{l|}{} & (1,$C$,$T$) \\ \cline{2-5}
 & \multicolumn{1}{|l|}{Conv2D} & \multicolumn{1}{l|}{$F_1$} & \multicolumn{1}{l|}{(1, $K_E$)} & \multirow{2}{*}{($F_1$,$C$,$T$)} \\ \cline{2-4}
 & \multicolumn{1}{|l|}{BatchNorm} & \multicolumn{1}{l|}{} & \multicolumn{1}{l|}{} &  \\ \hline
\multirow{5}{*}{$\phi^2$} & \multicolumn{1}{|l|}{DepthwiseConv2D} & \multicolumn{1}{l|}{$F_1 \cdot$2} & \multicolumn{1}{l|}{(C, 1)} & \multirow{3}{*}{(2$\cdot F_1$,1,$T$)} \\ \cline{2-4}
 & \multicolumn{1}{|l|}{BatchNorm} & \multicolumn{1}{l|}{} & \multicolumn{1}{l|}{} &  \\ \cline{2-4}
 & \multicolumn{1}{|l|}{EluAct} & \multicolumn{1}{l|}{} & \multicolumn{1}{l|}{} &  \\ \cline{2-5}
 & \multicolumn{1}{|l|}{AveragePool2D} & \multicolumn{1}{l|}{} & \multicolumn{1}{l|}{} & \multirow{2}{*}{(2$\cdot F_1$,1,$T$//8)} \\ \cline{2-4}
 & \multicolumn{1}{|l|}{Dropout} & \multicolumn{1}{l|}{} & \multicolumn{1}{l|}{} &  \\ \hline
\multirow{5}{*}{$\phi^3$} & \multicolumn{1}{|l|}{SeparableConv2D} & \multicolumn{1}{l|}{$F_2$} & \multicolumn{1}{|l|}{(1, 16)} & \multirow{3}{*}{($F_2$,1,$T$//8)}  \\ \cline{2-4}
 & \multicolumn{1}{|l|}{BatchNorm} & \multicolumn{1}{l|}{} & \multicolumn{1}{l|}{} &  \\ \cline{2-4}
 & \multicolumn{1}{|l|}{EluAct} & \multicolumn{1}{l|}{} & \multicolumn{1}{l|}{} &  \\ \cline{2-5}
 & \multicolumn{1}{|l|}{AveragePool2D} & \multicolumn{1}{l|}{} & \multicolumn{1}{l|}{} & \multirow{2}{*}{($F_2$,1,$T$//64)} \\ \cline{2-4}
 & \multicolumn{1}{|l|}{Dropout} & \multicolumn{1}{l|}{} & \multicolumn{1}{l|}{} &  \\ \hline
$\phi^4$ & \multicolumn{1}{|l|}{\textbf{TCN}} & \multicolumn{1}{l|}{$F_T$} & \multicolumn{1}{l|}{$K_T$} & $F_T$ \\ \hline
\multirow{2}{*}{$\phi^5$} & \multicolumn{1}{|l|}{Dense} & \multicolumn{1}{l|}{} & \multicolumn{1}{l|}{} & \multirow{2}{*}{4} \\ \cline{2-4}
 & \multicolumn{1}{|l|}{SoftMaxAct} & \multicolumn{1}{l|}{} & \multicolumn{1}{l|}{} & 
\end{tabular}
\begin{tablenotes}\footnotesize
\item[] $C$ = number of EEG channels, $T$ = number of time samples, $F_1$ = number of temporal filters, $F_2$ = number of spatial filters, $K_E$ = kernel size in first convolution, $K_T$ = kernel size in TCN module, and $F_T$ = number of filters in TCN module. For dropout in EEGNet inspired layers we use $p_e$, and in the TCN module we use $p_t$.
\end{tablenotes}
\end{threeparttable}
%}
\end{table}

\begin{table*}
\centering
\begin{threeparttable}
\caption{Classification accuracy (\%) and $\kappa$ scores on the 4-class MI BCI Competition IV-2a dataset.}
\label{tab:acc_and_kappa}
\begin{tabular}{cccccccccccc}
 & \multicolumn{6}{c|}{Fixed Networks} & \multicolumn{5}{c}{Variable Networks}  \\ %\cline{2-11}
 \multicolumn{1}{c|}{}& \multicolumn{2}{c|}{EEGNet\tnote{*} \cite{lawhern_eegnet_2018}} & \multicolumn{2}{c|}{Shallow ConvNet\tnote{*} \cite{schirrmeister_deep_2017}} & \multicolumn{2}{c|}{\textbf{\eegtcnet{}}} & \multicolumn{2}{c|}{EEGNet} & \multicolumn{2}{c|}{\textbf{\eegtcnet{}}} & \multicolumn{1}{c}{DFFN~\cite{li_densely_2019}} \\ %\cline{2-11}
 \multicolumn{1}{c|}{}& \multicolumn{1}{c}{\textit{Accuracy}} & \multicolumn{1}{c|}{\textit{$\kappa$}} & \multicolumn{1}{c}{\textit{Accuracy}}  & \multicolumn{1}{c|}{\textit{$\kappa$}}  & \textit{Accuracy} & \multicolumn{1}{c|}{\textit{$\kappa$}} & \textit{Accuracy} & \multicolumn{1}{c|}{\textit{$\kappa$}} & \textit{Accuracy} & \multicolumn{1}{c|}{\textit{$\kappa$}} & \textit{Accuracy} \\ \hline
\multicolumn{1}{c|}{\textbf{Subject 1}} & 84.34 & \multicolumn{1}{c|}{0.79} & \multicolumn{1}{c}{79.51} & \multicolumn{1}{c|}{0.73} & 85.77 & \multicolumn{1}{c|}{0.81} & 86.48 & \multicolumn{1}{c|}{0.82} & 89.32 & \multicolumn{1}{c|}{0.86} & 83.46 \\ 
\rowcolor[HTML]{dbdbdb}
\multicolumn{1}{c|}{\textbf{Subject 2}} & 54.06 & \multicolumn{1}{c|}{0.39} & \multicolumn{1}{c}{56.25} &\multicolumn{1}{c|}{0.42} & 65.02 & \multicolumn{1}{c|}{0.53} & 61.84 & \multicolumn{1}{c|}{0.49} & 72.44 & \multicolumn{1}{c|}{0.63} & 69.30  \\ 
\multicolumn{1}{c|}{\textbf{Subject 3}} & 87.54 & \multicolumn{1}{c|}{0.83} & 88.89 & \multicolumn{1}{c|}{0.85} & 94.51 & \multicolumn{1}{c|}{0.93} & 93.41 & \multicolumn{1}{c|}{0.91} & 97.44 & \multicolumn{1}{c|}{0.97} & 90.29  \\ 
\rowcolor[HTML]{dbdbdb}
\multicolumn{1}{c|}{\textbf{Subject 4}} & 63.59 & \multicolumn{1}{c|}{0.51} & \multicolumn{1}{c}{80.90} & \multicolumn{1}{c|}{0.75} & 64.91 & \multicolumn{1}{c|}{0.53} & 73.25 & \multicolumn{1}{c|}{0.64} & 75.87 & \multicolumn{1}{c|}{0.68} & 71.07  \\ 
\multicolumn{1}{c|}{\textbf{Subject 5}} & 67.39 & \multicolumn{1}{c|}{0.57} & 57.29 & \multicolumn{1}{c|}{0.43} & 75.36 & \multicolumn{1}{c|}{0.67} & 76.81 & \multicolumn{1}{c|}{0.69} & 83.69 & \multicolumn{1}{c|}{0.78} & 65.41  \\ 
\rowcolor[HTML]{dbdbdb}
\multicolumn{1}{c|}{\textbf{Subject 6}} & 54.88 & \multicolumn{1}{c|}{0.39} & 53.82 & \multicolumn{1}{c|}{0.38}  & 61.40 & \multicolumn{1}{c|}{0.49} & 59.07 & \multicolumn{1}{c|}{0.45} & 70.69 & \multicolumn{1}{c|}{0.61} & 69.45  \\ 
\multicolumn{1}{c|}{\textbf{Subject 7}} & 88.80 & \multicolumn{1}{c|}{0.85} & 91.67 & \multicolumn{1}{c|}{0.89}  & 87.36 & \multicolumn{1}{c|}{0.83} & 90.25 & \multicolumn{1}{c|}{0.87} & 93.14 & \multicolumn{1}{c|}{0.91} & 88.18  \\ 
\rowcolor[HTML]{dbdbdb}
\multicolumn{1}{c|}{\textbf{Subject 8}} & 76.75 & \multicolumn{1}{c|}{0.69} & 81.25 & \multicolumn{1}{c|}{0.75} & 83.76 & \multicolumn{1}{c|}{0.78} & 87.45 & \multicolumn{1}{c|}{0.83} & 86.71 & \multicolumn{1}{c|}{0.82} & 86.76  \\ 
\multicolumn{1}{c|}{\textbf{Subject 9}} & 74.24 & \multicolumn{1}{c|}{0.65} & 79.17 & \multicolumn{1}{c|}{0.72}  & 78.03 & \multicolumn{1}{c|}{0.71} & 82.95 & \multicolumn{1}{c|}{0.77} & 85.23 & \multicolumn{1}{c|}{0.80} & 93.54  \\ 
\hline
\multicolumn{1}{c|}{\textbf{Mean}}& 72.40 & \multicolumn{1}{c|}{0.63} & 74.31 & \multicolumn{1}{c|}{0.66}  & \textbf{77.35} & \multicolumn{1}{c|}{\textbf{0.70}} & 79.06 & \multicolumn{1}{c|}{0.72} & \textbf{83.84} & \multicolumn{1}{c|}{\textbf{0.78}} & 79.71  \\ 
\multicolumn{1}{c|}{\textbf{Std. Dev.}}& 13.27 & \multicolumn{1}{c|}{0.18} & 14.54 & \multicolumn{1}{c|}{0.19}  & \textbf{11.57} & \multicolumn{1}{c|}{\textbf{0.15}} & 12.28 & \multicolumn{1}{c|}{0.16} & \textbf{9.20} & \multicolumn{1}{c|}{\textbf{0.12}} & 10.79  \\ \hline
%\multicolumn{11}{l}{\scriptsize\textsuperscript{*}EEGNet model modified and reproduced}
\end{tabular} 
\begin{tablenotes}\footnotesize
\item[*]Reproduced
\end{tablenotes}

\end{threeparttable}
\end{table*} 

\subsubsection{Fixed \eegtcnet{}}
A global architecture method involves choosing global hyperparameters for all subjects, then training and testing each subject separately. 
A cross-validated grid search on the training set over the hyperparameters yields the following optimal architecture that performed best for all subjects: $F_1 =$ 8, $F_2 =$ 16, $K_E=$ 32, $K_T =$ 4, $L =$ 2, $F_T =$ 12, $p_e =$ 0.2, $p_t =$ 0.3, and with data standardization.
\subsubsection{Variable \eegtcnet{}}

% shown in Table~\ref{tab:variable_parameters},
%One architecture for all subjects approach involves choosing a global architecture then training and evaluating for each subject separately. 

%\subsection{Variable Networks}
The accuracy of most classifiers on the BCI Competition IV-2a highly varies among individual subjects, e.g., EEGNet achieves accuracies ranging from 54.06\%--88.80\%. 
%for subject 6 and 7, respectively.  
% 
This might originate from the rigid network and training structure, applying the same network as well as optimizing hyperparameters to all subjects. 
Hence, we propose to find optimal subject-specific network parameters (e.g., kernel size, number of filters, or use of data standardization) and training hyperparameters (e.g., dropout rate) of \eegtcnet{} using cross-validated grid search on the training for every individual subject. 
The test set is not touched for determining  the optimal parameters, thus keeping it compatible with the rules of the BCI Competition IV-2a. 
For comparison, the same procedure is applied to EEGNet.  

\subsection{Training Procedure}
Models were trained and tested in a Tensorflow environment on an NVIDIA GTX 1080 Ti GPU. 
The networks were developed with Keras.
We use the same training configuration when training the models proposed in this paper, where categorical cross-entropy loss is used, and the filter kernels are uniformly initialized following the procedure introduced in~\cite{he_deep_2016}.
The models are trained for 750 epochs with an Adam optimizer at a learning rate of 0.001 and a batch size of 64. 
These training hyperparameters are determined via cross-validation on the training set. 

% While training our \eegtcnet{} and EEGNet, both fixed and variable, we use the following training procedure to determine the optimal number of epochs:
% \begin{enumerate}
%     \item for each subject, train model on stratified 20-fold cross-validation,
%     \item find epoch where validation accuracy peaks per fold,
%     \item find the average epoch of where the accuracy peaks in each fold and train model on the full training set for as many epochs.
% \end{enumerate}
% The 20-fold cross-validation scheme is employed to try to mimic the effect an leave-one-out cross-validation scheme while still trying to keep computational time down.

\section{Experimental Results}
\subsection{Performance Metrics}
%\subsubsection{Environment}
%Models were trained and tested in a Tensorflow environment on a NVIDIA GTX 1080 Ti GPU. The networks were developed with Keras.
%The code, as well as trained models, are available\footnote{Will be open-sourced after blind review.}.
%\subsubsection{Performance Metrics}
% We validate the models according to the classification accuracy and Cohen's kappa value defined as:
% \begin{align*}
%     \mathrm{Accuracy} = \frac{\textnormal{TP}}{\textnormal{TP}+\textnormal{FN}}, \ \ \kappa = \frac{p_o - p_e}{1-p_e}
% \end{align*}
% where TP stands for true positive, FN stands for false negative, $p_o$ the observed agreement ratio (e.g., accuracy), and $p_e$ for the hypothetical probability of chance agreement or random classification rate. 
We evaluate the models according to the classification accuracy, which is the ratio between correctly classified trials and the total number of trials in the test set. 
Additionally, we report Cohen's $\kappa$-score defined as:
\begin{align*}
    \kappa = \frac{p_o - p_e}{1-p_e},
\end{align*}
where $p_o$ stands for the observed agreement ratio (e.g., accuracy) and $p_e$ for the hypothetical probability of chance agreement or random classification rate. 

We also report the number of parameters in each model and the number of MACs for inference. The calculation of MACs for a couple of different convolutional layers can be seen below:
\begin{align*}
    \textnormal{Conv2D} &= K_1 \cdot K_2 \cdot C_{in} \cdot C_{out} \cdot H_{out} \cdot W_{out}, \\
    \textnormal{Conv1D} &= K \cdot C_{in} \cdot C_{out} \cdot W_{out}, \\
    \textnormal{SeparableConv2D} &= (K_1 \cdot K_2 + C_{out}) \cdot C_{in} \cdot H_{out} \cdot W_{out},\\
    \textnormal{DepthWiseConv2D} &= K_1 \cdot K_2 \cdot C_{in}\cdot D \cdot H_{out} \cdot W_{out},%\\
   % \textnormal{Dense} &= \textnormal{Input neurons} \cdot \textnormal{Output neurons} 
\end{align*}
where $K$ stands for the kernel size, $C$ stands for the total number of channels. Then, the $H$ and $W$ are the height and width of the tensors, respectively.

Finally, we compare the memory footprint of the models during inference, which is defined here as the size of the two largest consecutive feature maps.
For the calculation of the memory size, we assume that both the feature maps and weights of all the networks can be quantized to 8 bits at negligible accuracy loss based on literature~\cite{schneider_2020}. % 

% The feature maps in the models we explore are of the following format: $(C,H,W)$, we then get the following formula:
% \begin{align}
% \label{eq:feature_map_size}
%     \textnormal{Size} = C_{k} \cdot H_{k} \cdot W_{k} + C_{k+1} \cdot H_{k+1} \cdot W_{k+1},
% \end{align}
% where the underscore indicates the two consecutive feature maps. 

\subsection{BCI Competition IV-2a}
% Structured approach:

% 1. Comparison of fixed networks, mainly work on Table III and also mention  that there is one better (TPCT)
Table~\ref{tab:acc_and_kappa} summarizes the accuracy and $\kappa$-scores for both fixed and variable EEGNet and \eegtcnet{}. 
Moreover, the table includes also the accuracy and $\kappa$-scores of the reproduced fixed Shallow ConvNet~\cite{schirrmeister_deep_2017}, and the accuracy of variable DFFN~\cite{li_densely_2019}, since detailed results were reported in the paper. 
By first comparing the fixed networks, \eegtcnet{} shows high robustness in classifying the nine subjects and achieves 77.35\% accuracy and a $\kappa$-score of 0.70. 
This is an increase of 4.95\% accuracy compared to EEGNet; moreover, the standard deviation of accuracy scores between subjects is 11.57\%, which is significantly lower than the one for EEGNet (13.27\%) and Shallow ConvNet (14.54\%).
% and an improvement of approximately $31.26\%$ compared to EEGNet. 
%
%It is also important to note that the best performing fixed model is TPCT, which has an accuracy score of 88.87\%. 

\begin{table*}
\caption{Optimal network hyperparameters for each subject of the BCI Competition IV-2a dataset.}
\label{tab:variable_parameters}
\resizebox{\textwidth}{!}{%
\begin{threeparttable}

\begin{tabular}{c|ccccccccc|ccccccccc}
\multicolumn{1}{c|}{\textbf{}} & \multicolumn{9}{c|}{\textbf{Variable \eegtcnet{}}} & \multicolumn{9}{c}{\textbf{Variable EEGNet}} \\
\multicolumn{1}{c|}{\textbf{Subject}} & \textit{1} & \textit{2} & \textit{3} & \textit{4} & \textit{5} & \textit{6} & \textit{7} & \textit{8} & \textit{9} & \textit{1} & \textit{2} & \textit{3} & \textit{4} & \textit{5} & \textit{6} & \textit{7} & \textit{8} & \textit{9} \\ \hline
\cellcolor[HTML]{dbdbdb}$K_T$ & \cellcolor[HTML]{dbdbdb}3 & \cellcolor[HTML]{dbdbdb}4 & \cellcolor[HTML]{dbdbdb}4 & \cellcolor[HTML]{dbdbdb}4 & \cellcolor[HTML]{dbdbdb}3 & \cellcolor[HTML]{dbdbdb}4 & \cellcolor[HTML]{dbdbdb}4 & \cellcolor[HTML]{dbdbdb}3 & \cellcolor[HTML]{dbdbdb}3 &  &  &  &  &  &  &  &  &  \\
$p_t$ & 0.3 & 0.2 & 0.3 & 0.2 & 0.2 & 0.3 & 0.3 & 0.3 & 0.2 &  &  &  &  &  &  &  &  &  \\
\cellcolor[HTML]{dbdbdb}$L$ & \cellcolor[HTML]{dbdbdb}3 & \cellcolor[HTML]{dbdbdb}2 & \cellcolor[HTML]{dbdbdb}2 & \cellcolor[HTML]{dbdbdb}3 & \cellcolor[HTML]{dbdbdb}4 & \cellcolor[HTML]{dbdbdb}3 & \cellcolor[HTML]{dbdbdb}2 & \cellcolor[HTML]{dbdbdb}3 & \cellcolor[HTML]{dbdbdb}4 &  &  &  &  &  &  &  &  &  \\
$F_T$ & 15 & 17 & 15 & 17 & 25 & 17 & 20 & 25 & 12 &  &  &  &  &  &  &  &  &  \\
\cellcolor[HTML]{dbdbdb}$F_1$ & \cellcolor[HTML]{dbdbdb}8 & \cellcolor[HTML]{dbdbdb}8 & \cellcolor[HTML]{dbdbdb}8 & \cellcolor[HTML]{dbdbdb}16 & \cellcolor[HTML]{dbdbdb}16 & \cellcolor[HTML]{dbdbdb}16 & \cellcolor[HTML]{dbdbdb}8 & \cellcolor[HTML]{dbdbdb}16 & \cellcolor[HTML]{dbdbdb}16 & \cellcolor[HTML]{dbdbdb}32 & \cellcolor[HTML]{dbdbdb}32 & \cellcolor[HTML]{dbdbdb}8 & \cellcolor[HTML]{dbdbdb}16 & \cellcolor[HTML]{dbdbdb}32 & \cellcolor[HTML]{dbdbdb}32 & \cellcolor[HTML]{dbdbdb}32 & \cellcolor[HTML]{dbdbdb}32 & \cellcolor[HTML]{dbdbdb}8 \\
$K_E$ & 32 & 64 & 64 & 32 & 64 & 32 & 32 & 64 & 64 & 128 & 128 & 64 & 32 & 32 & 64 & 32 & 32 & 64 \\
\cellcolor[HTML]{dbdbdb}$p_e$ & \cellcolor[HTML]{dbdbdb}0.2 & \cellcolor[HTML]{dbdbdb}0.2 & \cellcolor[HTML]{dbdbdb}0.2 & \cellcolor[HTML]{dbdbdb}0.1 & \cellcolor[HTML]{dbdbdb}0.2 & \cellcolor[HTML]{dbdbdb}0.1 & \cellcolor[HTML]{dbdbdb}0.1 & \cellcolor[HTML]{dbdbdb}0.2 & \cellcolor[HTML]{dbdbdb}0.2 & \cellcolor[HTML]{dbdbdb}0 & \cellcolor[HTML]{dbdbdb}0 & \cellcolor[HTML]{dbdbdb}0.1 & \cellcolor[HTML]{dbdbdb}0.1 & \cellcolor[HTML]{dbdbdb}0 & \cellcolor[HTML]{dbdbdb}0.2 & \cellcolor[HTML]{dbdbdb}0.2 & \cellcolor[HTML]{dbdbdb}0.2 & \cellcolor[HTML]{dbdbdb}0 \\
$S$ & True & False & True & True & True & True & True & True & True & False & False & False & False & False & False & False & False & False \\
\cellcolor[HTML]{dbdbdb}Parameters & \cellcolor[HTML]{dbdbdb}6144 & \cellcolor[HTML]{dbdbdb}6793 & \cellcolor[HTML]{dbdbdb}5815 & \cellcolor[HTML]{dbdbdb}12\,171 & \cellcolor[HTML]{dbdbdb}20\,526 & \cellcolor[HTML]{dbdbdb}12\,171 & \cellcolor[HTML]{dbdbdb}8184 & \cellcolor[HTML]{dbdbdb}16\,526 & \cellcolor[HTML]{dbdbdb}8176 & \cellcolor[HTML]{dbdbdb}15\,620 & \cellcolor[HTML]{dbdbdb}15\,620 & \cellcolor[HTML]{dbdbdb}2628 & \cellcolor[HTML]{dbdbdb}5252 & \cellcolor[HTML]{dbdbdb}12\,548 & \cellcolor[HTML]{dbdbdb}13\,572 & \cellcolor[HTML]{dbdbdb}12\,548 & \cellcolor[HTML]{dbdbdb}12\,548 & \cellcolor[HTML]{dbdbdb}2628
\end{tabular}

\begin{tablenotes}\footnotesize
\item[] $K_T$ : Kernel size in TCN module, $p_t$: Dropout rate in TCN module, $L$: \# of residual blocks, $S$: Standardize data  \item[] $F_T$: Filters in convolutional layers, $F_1$: Temporal filters in EEGNet, $K_E$: Kernel size in EEGNet, $p_e$: Dropout rate in EEGNet
\end{tablenotes}
\end{threeparttable}
}
\end{table*}

% 2. Going variable. State the gain for EEG-TCNet as well as EEGNet. Compare to other variable network DFFN. Also here, work mainly on Table III. Then at the end you can discuss Table II (still not fully sure how to put that).
%While training and evaluating EEG-TCNet, we saw that some hyperparameters worked better for some subjects that proved hard to classify, such as Subject~2 and Subject~6. We, therefore, trained a subject-specific or Variable network for both EEGNet and EEG-TCNet. 
When focusing on the variable networks in Table~\ref{tab:acc_and_kappa}, we see that the addition of subject-specific hyperparameters increased the performance of \eegtcnet{} by 6.49\%, achieving the highest accuracy of 83.84\%. 
%a staggering
Similarly, the introduction of variable hyperparameters in EEGNet improves the accuracy by 6.66\%. 
 Variable \eegtcnet{} outperforms variable DFFN by 4.13\%. 
 Again it is noteworthy that variable \eegtcnet{} exhibits the lowest standard deviation between subjects in both accuracy scores and $\kappa$-scores than other variable models.
 This underlines that our variable \eegtcnet{} not only improves on already well-performing subjects, but enables higher accuracy for otherwise poorly performing subjects, e.g., Subject~2, Subject~5, or Subject~6. 
 
 % The improvement that EEG-TCNet delivers compared to EEGNet is roughly $40.34\%$. 

Table~\ref{tab:variable_parameters} summarizes the optimal subject-specific parameters for variable \eegtcnet{} and EEGNet. 
Except for Subject~2, variable \eegtcnet{} always makes use of data standardization, whereas variable EEGNet consistently classifies the raw data without standardization. 
Interestingly, variable \eegtcnet{} requires, in general, a smaller number of temporal filters $F_1$ and filter size $K_E$ than EEGNet.
The temporal filters pose the most restrictive limitations in terms of computational complexity and memory footprint since the temporal convolution requires the vast majority of MACs and memory to store the resulting feature maps~\cite{wang_accurate_2020}.  
Therefore, variable \eegtcnet{} has more potential to be embedded on a resource-limited device. 
%
%The dropout and number of residual blocks in the TCN module were optimal for each subject at the rate of 0.3 and two to three blocks, respectively.

%The variable hyperparameters can then be seen by looking at Table~\ref{tab:variable_parameters}, we see two hyperparameters that are fixed or almost fixed for each subject in the Variable EEG-TCNet. The dropout and number of residual blocks in the TCN module were optimal for each subject at the rate of 0.3 and two to three blocks, respectively. The data standardization, $S$, hyperparameter is almost the same for each subject, varying only in Subject~2. The kernel size $K_E$ varies between $32$ and $64$ and no apparent connection between the kernel size in the Variable EEG-TCNet and EEGNet. Other hyperparameters vary much more between subjects, and no obvious connections can be seen between hyperparameters found in the Variable EEG-TCNet and EEGNet. The hyperparameters in Variable EEGNet also vary much between subjects, with the exception that it performed best without data standardization.

% 3. Now we compare the networks in terms of number of parameters, MACs and feature maps, referring to table IV
\begin{table}
\caption{Metrics of current SoA models of BCI Competition IV-2a.}
\label{tab:accuracies}
\resizebox{\columnwidth}{!}{%
\begin{tabular}{lrrrr}
 & \textbf{\begin{tabular}[c]{@{}c@{}}Mean\\ Accuracy\end{tabular}} & \multicolumn{1}{c}{\textbf{Parameters}} & \textbf{\begin{tabular}[c]{@{}c@{}}Mean\\ MACs\end{tabular}}  & \textbf{\begin{tabular}[c]{@{}c@{}}Feature\\ Map [kB]\end{tabular}}\\ \hline
\multicolumn{1}{l|}{EEGNet\rlap{\textsuperscript{*}}~\cite{lawhern_eegnet_2018}} & 72.40 & 2.63\,k & 13.1\,M & 396 \\
\multicolumn{1}{l|}{Shallow ConvNet\rlap{\textsuperscript{*}}~\cite{schirrmeister_deep_2017}} & 74.31 & 47.3\,k & 63.0\,M & 1013 \\
\multicolumn{1}{l|}{FBCSP~\cite{hersche_fast_2018}} & 73.70 & 261\,k & 104\,M & 50 \\
\multicolumn{1}{l|}{Riemannian~\cite{hersche_fast_2018}} & 74.77 & 50.0\,k & - & 49 \\
\multicolumn{1}{l|}{MSFBCNN~\cite{wu_parallel_2019}} & 75.80 & 155\,k & 202\,M & 5775 \\
\multicolumn{1}{l|}{\textbf{\eegtcnet{}}} & \textbf{77.34} & \textbf{4.27\,k} & \textbf{6.8\,M} & \textbf{396} \\
\multicolumn{1}{l|}{CNN++~\cite{zhao_improvement_2017}} & 81.10 & 240\,k & 96.4\,M & 499\\
\multicolumn{1}{l|}{TPCT~\cite{li_novel_2020}} & 88.87 & 7.78\,M & 1.73\,G & 524\\
\rowcolor[HTML]{dbdbdb}
\multicolumn{1}{l|}{\textit{Variable} EEGNet} & 79.02 & 15.6\,k & 42.6\,M & 1584\\
\rowcolor[HTML]{dbdbdb}
\multicolumn{1}{l|}{DFFN (variable)~\cite{li_densely_2019}} & 79.71 & 1.07\,M & 132\,M & 650\\ 
\rowcolor[HTML]{dbdbdb}
\multicolumn{1}{l|}{\textit{\textbf{Variable \eegtcnet{}}}} & \textbf{83.84} & \textbf{20.5\,k} & \textbf{12.1\,M} & \textbf{792}\\
\multicolumn{5}{l}{\scriptsize\textsuperscript{*}Reproduced}%
\end{tabular}
}
\end{table}

Table~\ref{tab:accuracies} compares the current SoA networks on the BCI Competition IV-2a in terms of accuracy, the number of trainable parameters, MACs, and memory requirements. 
For the variable models, we report the maximum number of parameters and memory requirements, as they pose a hard requirement when considering the embedding to an MCU. 
%
%By first looking at the parameter count and MACs of fixed networks, we see that the high increase of both accuracy and $\kappa$-scores of \eegtcnet{}, from the plain EEGNet, comes at a negligible increase of 40\,000 MACs. 
By first looking at the parameter count and MACs of fixed networks, we see that \eegtcnet{} requires only 6.8\,MACs, which is 1.9$\times$ lower than EEG-Net, while being 4.95\% more accurate. 
The large reduction in complexity comes from the smaller temporal filter size, where fixed \eegtcnet{} uses $K_T=$32 instead of $K_T=$64.
When allowing subject-specific network hyperparameters in the variable \eegtcnet{}, the maximum number of parameters increases by 4.80$\times$ and the MACs by 1.78$\times$, compared to fixed \mbox{\eegtcnet{}}. 
TPCT achieves the highest accuracy of 88.87\%. 
However, this comes at the cost of a 380$\times$ higher number of parameters and 143$\times$ more MACs than variable \eegtcnet{}. 

%
%TPCT achieves results at the cost of an immensely high parameter count and very costly inference cost. 
%Then by looking at the variable models, it is also noteworthy that EEG-TCNet delivers excellent performance with around 90 times fewer parameters and around 8-times fewer MACs than DFFN. 
%

Another consideration is the maximum memory requirements of the networks, which---assuming layer-by-layer inference---is the sum of the two largest consecutive feature maps.
Due to their residual structure, DFFN~\cite{li_densely_2019} and MSFBCNN~\cite{wu_parallel_2019} are calculated differently; DFFN needs to store the first five feature maps in the first dense block, and MSFBCNN~\cite{wu_parallel_2019} the first five feature maps as later layers depend on all of these feature maps rather than the immediately  preceding layer's output.
This results in MSFBCNN having the largest feature map size while DFFN manages to keep it relatively low despite the network architecture. 
Interestingly, TCPT has the fourth-smallest feature map size despite having the largest parameter count and most number of MACs. 

TPCT's overall memory footprint is 8.304\,MB, and thus far beyond the on-chip memory capacity available in an ARM M7 processor.
Its compute effort of 1.73\,GMACs would take approximately 50\,s/inference---17$\times$ below real-time when requiring a new classification at least every 3\,s---where we refer to the throughput of 34.45\,MMAC/s of an ARM M7 processor~\cite{wang_accurate_2020}. 
In comparison, the proposed EEG-TCNet has a memory footprint of 400\,kB, and its compute effort of 6.8\,MMACs would take approximately 197\,ms. 
% where we refer to the performance of the ARM M7 processor~\cite{wang_accurate_2020}.
%
\mbox{\eegtcnet{}} and variable \eegtcnet{} are the best candidates for an embedded implementation; both parameter count and inference cost are kept reasonable while still achieving very high accuracy scores.

\begin{figure*} 
\fontsize{7}{10}\selectfont
    \centering
  \subfloat[Accuracy vs. parameters, size of circles $\propto$ MACs.\label{fig:accvspara}]{%
       \includesvg[width = 0.95\columnwidth]{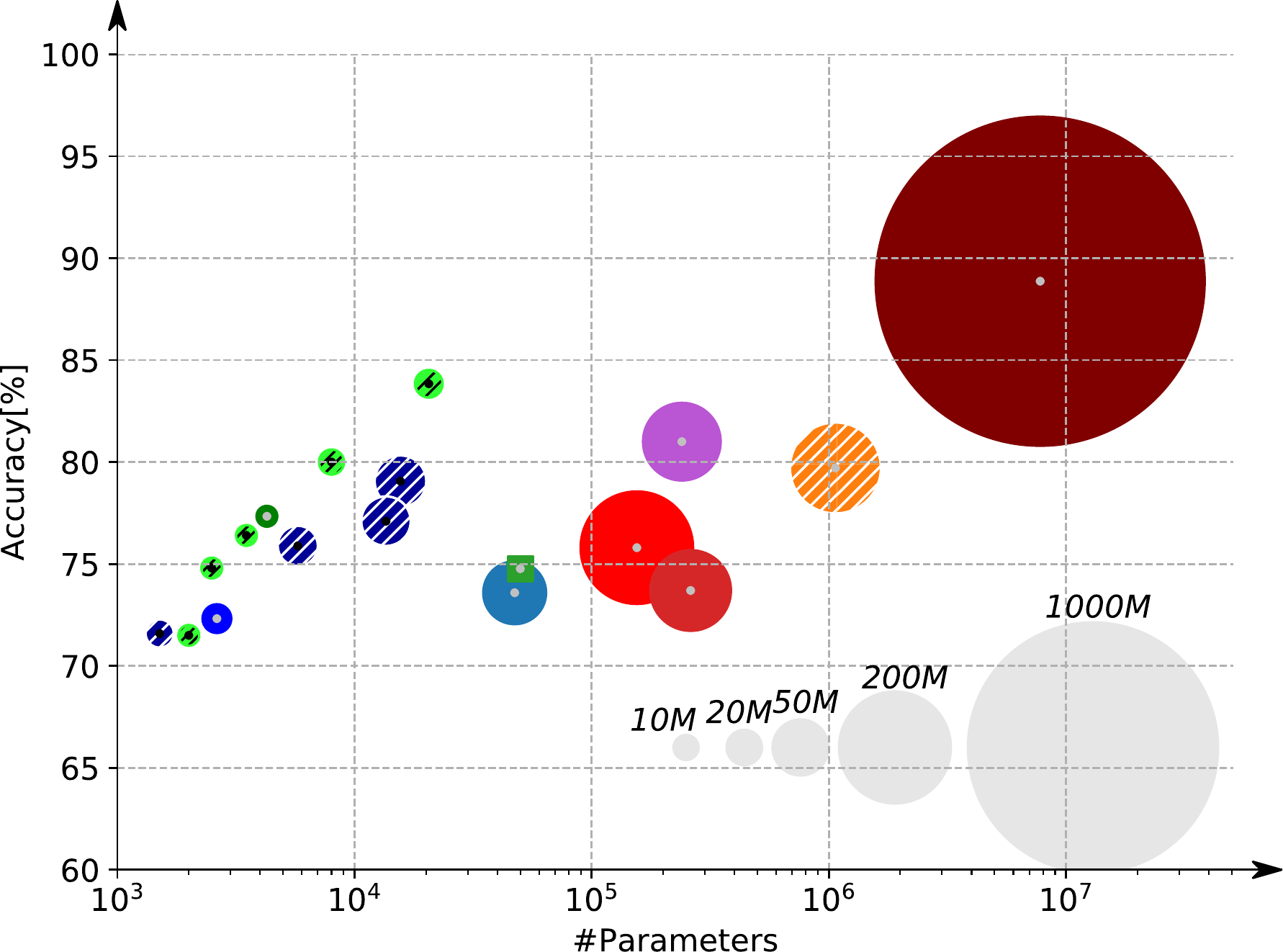}}
    \hfill
  \subfloat[Accuracy vs. MACs, size of circles $\propto$ number of parameters.\label{fig:accvsmacs}]{%
        \includesvg[width = 0.95\columnwidth]{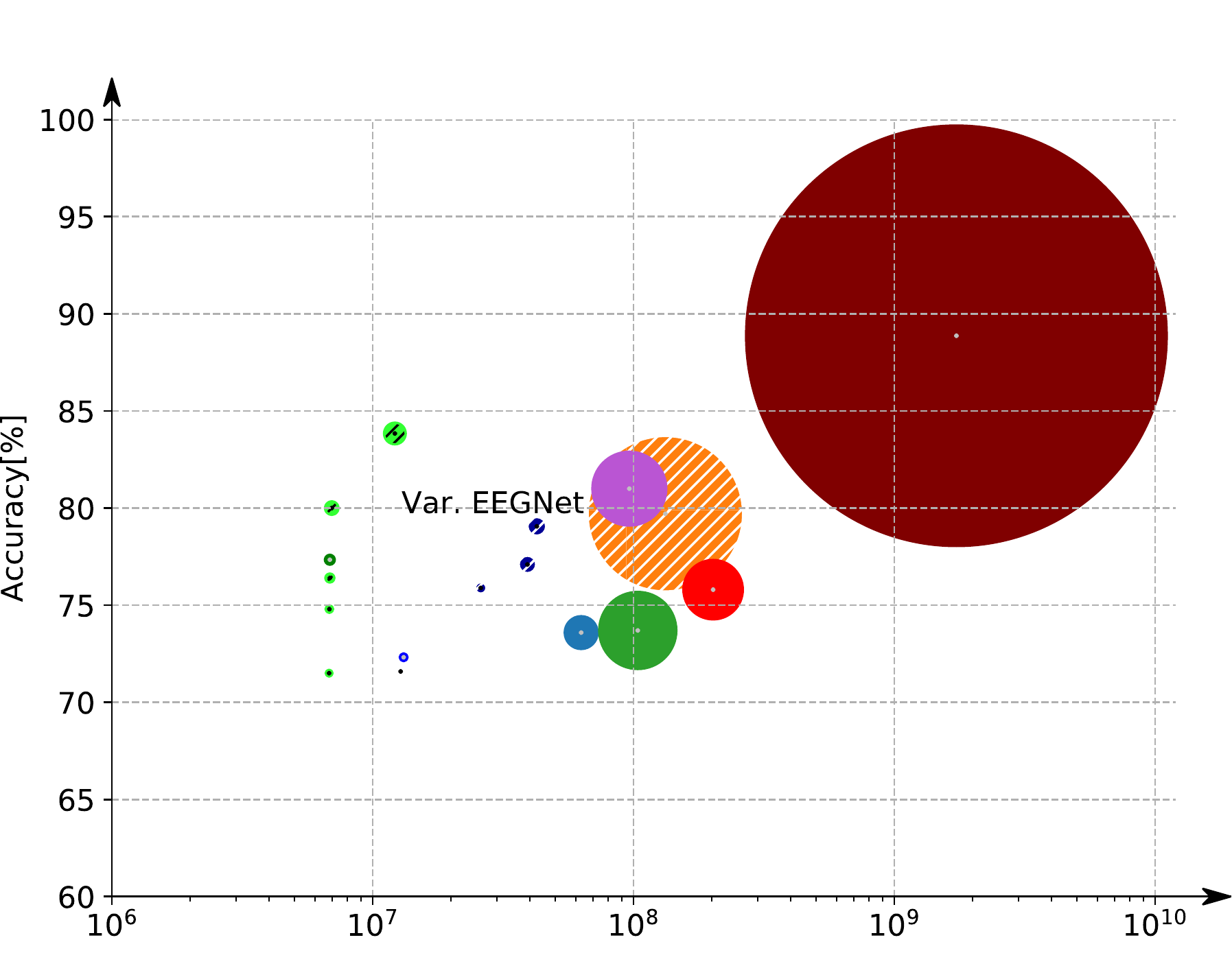}}
  \caption{Classification accuracy on BCI Competition IV-2a vs. (a) parameters and (b) MACs per inference.
  The circles are proportional to (a) MACs and (b) parameters, with the corresponding legend in grey reported on the bottom right.
The grey dots inside the circles highlight their center. % of the circles. 
%   %
The hatched circles show variable subject-specific networks. }
  \label{fig:accvspara_mac} 
\end{figure*}

Fig.~\ref{fig:accvspara_mac} visualizes the trade-off between the accuracy, number of parameters, and MACs of all models. 
We also include FBCSP and Riemannian~\cite{hersche_fast_2018}; the Riemannian model has a square shape in Fig.~\ref{fig:accvspara} and is not included in Fig.~\ref{fig:accvsmacs} as the number of MACs is not clear as its compute workload is not predominantly based on MAC operations. 
We further experiment with limiting the number of parameters that the variable models are allowed to have in each network. % and looked at the accuracy scores after choosing models that were below the limit. 
Specifically, we limit \eegtcnet{} to 2\,k, 2.5\,k, 3.5\,k, and 8\,k parameters and EEGNet to 1.5\,k, 5.7\,k, and 13.5\,k. 
%
%These limits and accuracy scores can be seen in Fig.~\ref{fig:accvspara}. 
We recognize that \eegtcnet{} achieves Pareto optimality by spanning almost the entire Pareto front in both the parameter and MAC comparison.
\subsection{Mother of All BCI Benchmarks}
\begin{figure}
    \fontsize{8}{10}\selectfont
    \centering
    \includesvg[width=\columnwidth]{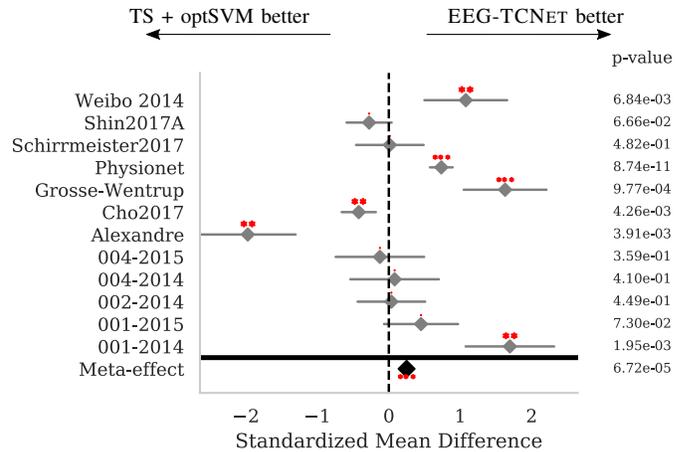}
    \caption{Meta-analysis comparing tangent space features on optimized SVM (TS + optSVM) against \eegtcnet{} on MOABB. The effect sizes shown are standardized mean differences, with p-values corresponding to the one-tailed Wilcoxon signed-rank test for the hypothesis given at the top of the plot and 95\% interval denoted by the grey bar. Stars correspond to *** =$p <$ 0.001,
    ** $= p <$ 0.01, * $= p <$ 0.05. The meta-effect is shown at the bottom.}
 \label{fig:meta_analysis}
\end{figure}

%About MOABB results
%Three other pipelines were tested against \emph{EEG-TCNet}, note these are three of the pipelines used and tested in the MOABB paper~\cite{jayaram_moabb_2018}.
We benchmark \eegtcnet{} and EEGNet on MOABB by comparing them to three other pipelines included in MOABB~\cite{jayaram_moabb_2018}.
These pipelines are: 
\begin{itemize}
\item CSP + LDA: where trial covariances were estimated via maximum-likelihood with unregularized CSP.
Features were log-variance of the filters belonging to the six most diverging eigenvalues and then classified with LDA.
\item TS + optSVM: where trial covariances were estimated via oracle approximating shrinkage, then projected into the Riemannian tangent space to obtain features and classified with a linear SVM with identical grid search.
\item AM + optSVM: where features are the log-variance in each channel and then classified with a linear SVM with grid search.
\end{itemize}
%
%At the end of the MOABB procedure, the scores of every subject in every dataset are stored, and statistical analysis is done. 
Fig.~\ref{fig:meta_analysis} shows the meta-analysis of the comparison between \eegtcnet{} and TS + optSVM, which is the current SoA on MOABB. 
% using Riemannian tangent space features together with an optimized SVM. 
%
The meta-effect reports the combined standardized mean differences across all datasets. The standardized mean difference is combined with a weighting given by the square root of the number of subjects. Then, the p-value of a one-tailed Wilcoxon signed-rank test for the hypothesis that \mbox{\eegtcnet{}} is more accurate is also reported. 
%
%To get a statistical overview of how the \emph{EEG-TCNet} compares to other algorithms, a meta-analysis is done in Fig.~\ref{fig:meta_analysis}.
%
While we observe a high variance among different datasets that could give contradictory results if the methods were evaluated on one dataset in isolation, the overall trend shows that \eegtcnet{} outperforms TS + optSVM with a final meta-effect of 0.25. 

\begin{figure}
    \fontsize{8}{10}\selectfont
    \centering
    \includesvg[width=0.9\columnwidth]{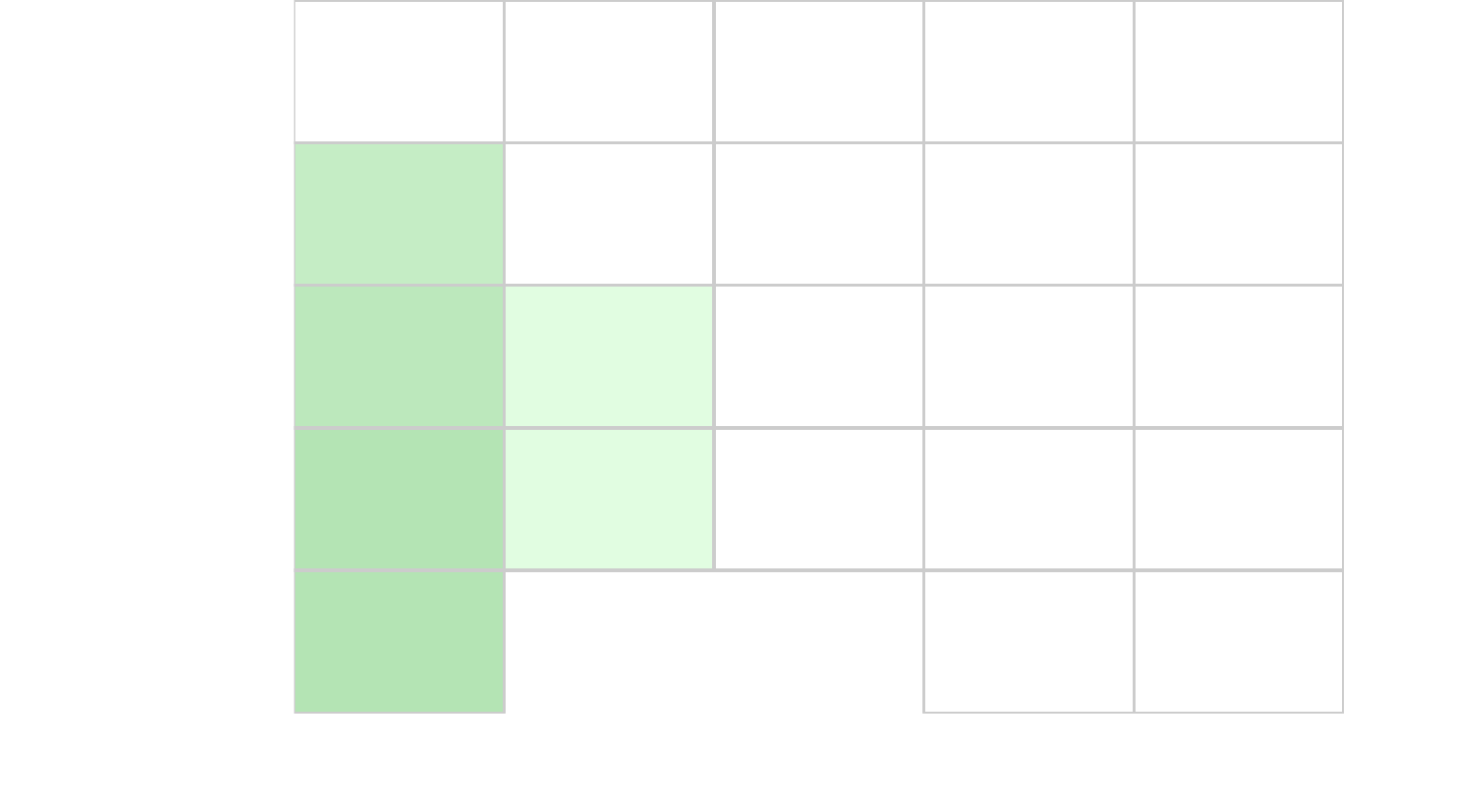}
    \caption{
    %Ranking of algorithms in performance across all datasets. 
    Meta-effect on MOABB with the hypothesis that the method on the vertical axis performs better then the one on the horizontal axis.
    All p-values are single-sided; in the case the effect goes in the opposite direction of the hypothesis, the values are removed for clarity. 
    The values correspond to the standardized mean difference of the algorithm in the y-axis minus that in the x-axis and the associated p-value.}
    \label{fig:MOABB_ordering}
\end{figure}

Fig.~\ref{fig:MOABB_ordering} summarizes the comparisons between all methods, showing the meta-effect in case that the method on the vertical axis significantly outperforms the method on the horizontal axis, according to the one-tailed Wilcoxon signed-rank test. 
\eegtcnet{} outperforms all other methods; thus, it is becoming the new SoA on MOABB. 
This experiment underlines that \eegtcnet{} generalizes well outside a single MI dataset, where it was modeled on. 
%
%Even though \emph{EEG-TCNet} is modeled after a single dataset, it is still representative under the very robust validation framework that MOABB provides.

\section{Conclusion}
We have proposed \eegtcnet{}, a novel model for accurate MI-BMI classification. Thanks to its low memory footprint and limited computational complexity, it can be easily operated on low-power resource-limited devices at the edge. %
It achieves 77.35\% accuracy on the BCI Competition IV-2a dataset, improving the SoA of similarly-sized networks by 4.95\%. 
We then further enhance the model by performing a subject-specific hyperparameter search, which yields an additional 6.49\% accuracy increase, achieving a high accuracy of 83.84\%. 
Moreover, it requires a low number of parameters, MACs, and memory usage during inference. 
%
%Next to explore if \emph{EEG-TCNet}, which was designed and modeled for one BCI-MI dataset, was able to generalize outside that dataset, it was ported over to the MOABB framework and put under robust validation. 
%
Large scale benchmark tests on MOABB confirm that \eegtcnet{} generalizes well to other MI datasets, becoming the new SoA on the MOABB framework outperforming the old SoA by a final meta-effect of 0.25. 
%

%Which makes it an exciting and light model that can then be ported over to embedded platforms for the intended use of BCI-MI classification tasks.

% Print the bibliography.
%\nocite*{} % Print non-cited references as well.
\bibliographystyle{IEEEtran}
\bibliography{misc.bib}%Bib/bibliography.bib}%,Bib/michael_mendeley.bib}

\end{document}